\documentclass[12pt]{article}
\usepackage{epsf}
\usepackage{amsmath}
\def\be{\begin{equation}}
\def\ee{\end{equation}}

\def\pp{\psi(2S)}
\def\jp{J/\psi}

\begin{document}
\begin{titlepage}
\begin{center}
{\Large \bf William I. Fine Theoretical Physics Institute \\
University of Minnesota \\}
\end{center}
\vspace{0.2in}
\begin{flushright}
ITEP-TH-18/08 \\
FTPI-MINN-08/12 \\
UMN-TH-2643/08 \\
April 2008 \\
\end{flushright}
\vspace{0.3in}
\begin{center}
{\Large \bf Holographic Hadro-Quarkonium
\\}
\vspace{0.2in}
{\bf S. Dubynskiy \\}
School of Physics and Astronomy, University of Minnesota, \\ Minneapolis, MN
55455, USA, \\
{\bf A. Gorsky \\}
Institute of Theoretical and Experimental Physics, Moscow, 117218, Russia \\
and \\
{\bf M.B. Voloshin  \\ }
William I. Fine Theoretical Physics Institute, University of
Minnesota,\\ Minneapolis, MN 55455, USA \\
and \\
Institute of Theoretical and Experimental Physics, Moscow, 117218, Russia
\\[0.2in]
\end{center}

\begin{abstract}

We consider the recently suggested model for some resonances near the open charm
threshold as bound states of charmonium inside excited light mesons. It is
argued in the soft-wall holographic model of QCD  that such states of heavy
quarkonium necessarily  exist at sufficiently large spin of the light meson. The
bound state is provided by the dilaton  exchange  through the 5D bulk. We also
argue that the decay of such bound systems into mesons with open heavy flavors
due to splitting of the heavy quarkonium can be treated as semiclassical
tunneling and is suppressed. This behavior is in agreement with the known
relative suppression of the decay of the discussed charmonium-like resonances
into channels with $D$ mesons.

\end{abstract}

\end{titlepage}

\section{Introduction}
Recent observation\footnote{Recent reviews can be found in
Refs.~\cite{egmr,mvc,go}.} of resonances near the open charm threshold which
decay into a specific charmonium state, $\jp$ or $\pp$, and two or even one pion
with no other observed conspicuous decay channels\,\cite{noother}
suggests\,\cite{mvc} that these resonances in fact contain the specific
charmonium
state `coated' by an excited light-hadron matter. It has been further
argued\,\cite{dv} that such hadro-charmonium systems are likely to arise due to
binding of charmonium inside sufficiently excited light resonances through a QCD
analog of the van der Waals interaction.

In this paper we investigate in some detail the interaction of heavy quarkonium
with excited light mesons within the simplest soft wall model of holographic
QCD\,\cite{ekss} which provides linear behavior of Regge
trajectories\,\cite{kkss,metsaev} and a reasonable behavior of a heavy-quark
potential\,\cite{az}. In this approach the mesons can be considered as modes of
the flavor gauge field in 5D theory (vector mesons) or a Wilson line of this
gauge field along the fifth coordinate ($\pi$-meson). Large-spin mesons can be
considered within this approach as well. We also have to introduce heavy
quarkonium in this framework. It is clear that  the corresponding heavy degrees
of freedom are localized near the boundary at $z=0$ at scales of order $M_Q^{-
1}$, where $M_Q$ is the mass of the heavy quark. In the leading approximation we
shall substitute the heavy meson by the point-like source. The essential local
operator on the boundary can be deduced from the physical arguments  and is
approximated by the trace of the stress tensor. Since this operator interacts
with the dilaton we have a point-like dilaton source in the 5D theory. On the
other hand the light exited meson with large spin is mainly localized on the IR
wall scale\,\cite{son1}. Although it is better described by a long open string,
this approximation is also reliable. Thus the appropriate configuration consists
of a local source at the boundary $z=0$  and extended meson at the $z_{IR}$
interacting via dilaton exchange. It is shown that the interaction is strong
enough to yield the bound state at a sufficiently large spin $S$ of the light
meson.

Another feature which has to be explained is the phenomenologically apparent
suppression of the decays of hadro-quarkonium into states with pairs of heavy
mesons with open flavor. In the discussed picture such decays would correspond
to dissociation of heavy quarkonium due to forces from light degrees of
freedom. We argue that in the limit of large heavy quark mass $M_Q$ such
dissociation can be described by semiclassical tunneling and is suppressed by
the factor $\exp(-const \cdot M_Q^{1/2}/\Lambda_{QCD}^{1/2})$. Within the
holographic approach this process is a  generalization of the picture discussed
in \cite{son2}. Clearly, the suppression of the dissociative decays is in
qualitative agreement with the available observations.

The subsequent material in the paper is organized as follows. In Sec.\,2 a
description of the interaction of heavy quarkonium with soft gluonic field is
recapitulated and formulated in terms of the trace of the stress
tensor in low-energy QCD. In Sec.\,3 the interaction of the heavy quarkonium
with light meson is described in terms of the soft wall holographic model and in
Sec.\,4 it is shown that this interaction necessarily results in existence of
bound states for sufficiently high orbital excitation of the light meson. In
Sec.\,5 we consider the dissociation process for decays of the bound states into
heavy meson pairs with open flavor and argue that such process generally carries
a semiclassical suppression at large mass of the heavy quark. Finally, in
Sec.\,6 we summarize our conclusions.

\section{Quarkonium interaction with gluons}

Due to the binding between heavy quark and antiquark, the states of heavy
quarkonium are relatively compact in the standard hadronic scale set by
$\Lambda_{QCD}$. For this reason the interaction of quarkonium with gluonic
fields inside a light meson can be considered in terms of multipole
expansion\,\cite{gottfried,mv79}. The leading nonrelativistic (in the heavy
quarks) contribution to the van der Waals type interaction then arises in the
second order in the chromoelectric dipole term, and can be described by the
effective Hamiltonian\,\cite{peskin,mvpol}
\be
H_{\rm eff}= -{1 \over 2} \, \alpha^{(Q \bar Q)} \, E_i^a E_i^a~,
\label{hdiag}
\ee
where ${\vec E}^a$ is the chromoelectric field and $\alpha^{(Q \bar Q)}$ is the
chromo-polarizability, depending on particular state of the $(Q \bar Q)$ system,
in complete analogy with the description in terms of polarizability of the
interaction of atoms with long-wave electric field.

The values of the chromo-polarizability for charmonium and bottomonium levels
below the open flavor threshold are all real and positive\,\cite{mvpol,sv}. The
numerical values are so far known only for the off-diagonal polarizability,
describing the observed $\pi \pi$ transitions in charmonium and bottomonium. The
`reference' values are\,\cite{mvpol} $\alpha^{(\jp \, \psi')} \approx
2\,$GeV$^{-3}$ in charmonium and $\alpha^{(\Upsilon \Upsilon')}\approx
0.6\,$GeV$^{-3}$ in bottomonium. It is fully expected\,\cite{mvpol,sv} that
proper
diagonal values of the chromo-polarizability in both systems are larger than
these `reference' values, especially for excited states. The quoted values
correspond to the normalization convention for the gluonic field strength with
the QCD coupling $g$ included in the field strength $F_{\mu \nu}^a$.

In what follows we consider the interaction (\ref{hdiag}) with the operator
${\vec E}^a \cdot {\vec E}^a$ replaced by $-(F_{\mu \nu}^a)^2/2 =  {\vec E}^a
\cdot {\vec E}^a - {\vec B}^a \cdot {\vec B}^a$, with ${\vec B}^a$ being the
chromomagnetic field. In doing so we consider attraction of quarkonium into
gluonic medium which is weaker by the contribution of the manifestly
sign-definite term ${\vec B}^a \cdot {\vec B}^a$ than the actual interaction
described by the Hamiltonian (\ref{hdiag}). This clearly leads to a conservative
treatment of the problem of existence of bound states\,\cite{sv,dv}.
Furthermore,
the operator $(F_{\mu \nu}^a)^2$ is related to the trace of the stress tensor
$\theta_\mu^\mu$ in QCD with (three) light quarks in the chiral limit by the
well known anomaly relation:
\be
\theta_\mu^\mu={\beta(g^2) \over 4 \, g^4}\, (F_{\mu \nu}^a)^2 \approx - \, {9
\over 32 \pi^2} \, (F_{\mu \nu}^a)^2~,
\label{anom}
\ee
where we replace the QCD beta function by its first one-loop term for
definiteness of subsequent numerical estimates, while our general conclusion
about existence of bound states does not depend on this replacement.
Therefore the form of the van der Waals interaction of heavy quarkonium with
light hadronic matter that we consider here is
\be
H_W=-C \, \theta_\mu^\mu~
\label{hw}
\ee
with $C=(8 \, \pi^2/ 9) \, \alpha^{(Q \bar Q)} \, [1 +O(g^2)]$.

We consider the limit of large $M_Q$ so that the kinetic energy of the motion of
heavy quarkonium is totally neglected, and we choose the position of quarkonium
(in the three dimensional space) at the origin $\vec x =0$.
In a holographic model heavy quarkonium is localized at $z=0$, so that the
interaction (\ref{hw}) corresponds to a localized at $(z=0, \vec x=0)$ source of
dilaton.

\section{Holographic interaction of light mesons with heavy quarkonium}

In a holographic soft wall model\,\cite{kkss} an orbitally excited light meson
with spin $S$ is described by normalizable solutions $\phi(z,x^\mu)$ of the
five-dimensional equation:
\be
\left ( \partial_\mu \partial^\mu - {\partial^2 \over \partial z^2} + z^2 + 2\,
S -2 + {S^2 - 1/4 \over z^2} \right ) \, \phi(z,x) =0~.
\label{5deq}
\ee
Upon substitution of a four-dimensional plane wave $\phi(z,x) = e^{-i p \cdot x}
\, \psi(z)$ this results in the eigenvalue problem for the masses $m^2 = p^2$ of
the mesons:
\be
\left (  - {{\rm d}^2 \over {\rm d}z^2} + z^2 + 2\, S -2 + {S^2 - 1/4 \over z^2}
\right ) \, \psi_n(z) = m_n^2 \, \psi_n(z)~,
\label{zeig}
\ee
where $n$ is the excitation number at a given $S$. The spectrum of $m_n^2$ is
then directly found from the obvious relation of this problem to a
two-dimensional harmonic oscillator, and is given by\,\cite{kkss}
\be
m_n^2 = 4(n+S)~,
\label{rspec}
\ee
thus reproducing linear Regge trajectories.

It can be mentioned that here dimensionless units are used, corresponding to the
coefficient of the ``soft wall" term $z^2$ equal to one. In normal units the
energy scale is set by a coefficient $\sigma$ proportional to the string
tension.

In the presence of a static source of the dilaton field located at $(z=0, \vec x
=0)$ the equation (\ref{5deq}) is generally no longer translationally invariant
in $\vec x$, and the problem becomes a (3+1) dimensional eigenvalue problem for
the energy $\omega$:
\be
\left [ -\Delta_x - {\partial^2 \over \partial z^2} + z^2 + 2\, S -2 + {S^2 -
1/4 \over z^2} + V(z, \vec x)  \right ] \, \chi_n(z,\vec x) = \omega_n^2 \,
\chi_n(z,\vec x)~,
\label{5deqm}
\ee
where $\Delta_x$ is the three-dimensional Laplacian and $V(z, \vec x)$ is the
potential resulting from the dilaton propagation from the source and interaction
with the light-hadron string:
\be
V(z, \vec x) = g(z)\, D(z,{\vec x}) \, \eta~,
\label{potgen}
\ee
with $D(z, \vec x)$ being the dilaton propgator from the boundary ($z=0$, $\vec 
x=0$) to the bulk ($z,\, \vec x$), integrated over time (static source), $\eta$ 
is the strength of the source determined by Eq.(\ref{hw}), and $g(z)$ is the 
vertex for the dilaton-string interaction. The $z$ dependence of the latter 
vertex arises due to the assumed in Eq.(\ref{5deq}) $z$ dependent conformal 
symmetry breaking term $z^2$.

We will show that for the purpose of present discussion there is in fact no need
to determine each of the factors in Eq.(\ref{potgen}) separately, and that the
potential $V(z, \vec x)$ can be found using the normalization condition for the
stress tensor and the fact that the propagation in the bulk of the (conformal
dimension $\Delta=4$) dilaton field from a static source is described by the
operator
\be
-\Delta_x - {\partial^2 \over \partial z^2} + z^2 + 2 + {15 \over 4 \, z^2}~.
\label{dilop}
\ee
The $z$ part of this operator corresponds to the motion of two-dimensional
harmonic oscillator with angular momentum $j=2$, and the spatial (${\vec x}$)
part corresponds to a free motion. Thus one can readily find the propagator as
an integral over the proper time $\tau$ of the corresponding evolution
kernel\,\cite{fh} and write
\be
V(z, \vec x)  = f(z) \int_0^\infty \, {\rm d} \tau \, \left ( {1 \over 2 \pi
\tau} \right )^{3/2} \, \exp \left(-  {x^2 \over 2\tau} \right ) \, { e^{-\tau}
\over \sinh^3 \tau} \, \exp \left ( - {z^2 \over 2}\, {\cosh \tau \over  \sinh
\tau} \right )~.
\label{pot1}
\ee
The function $f(z)$ here includes the vertex factor $g(z)$ and the source
strength $\eta$ from Eq.(\ref{potgen}) as well as normalization and
$z$-dependent metric factors in the propagator $D(z,{\vec x})$.

The function $f(z)$ can be found by comparing the matrix elements of the
Hamiltonian (\ref{hw}) and the potential (\ref{pot1}) over the plane-wave in $x$
solutions of the equation (\ref{5deq}). Indeed for {\it any} eigenstate in
Eq.(\ref{zeig}) the normalization of the trace of the stress tensor requires
\be
\langle \phi_n(z,x) \, | \, \theta_\mu^\mu \, | \, \phi_n(z,x) \rangle = 2 \,
m_n^2 \, \langle
\phi_n(z,x) \, | \, \phi_n(z,x) \rangle~,
\label{norm}
\ee
which determines the average of the interaction (\ref{hw}) over those states. On
the other hand,  the average of the potential over the same states is given by 
its integral over $\vec x$:
\be
\langle \phi_n(z,x) \, | \, V(z, \vec x) \, | \, \phi_n(z,x) \rangle = \langle
\psi_n(z) \, | \,
\int V(z, \vec x) \, {\rm d}^3 x \, | \, \psi_n(z) \rangle = \langle \psi_n(z)\,
|\, f(z)
\, {4 \, e^{- z^2 / 2} \over z^4} \, | \, \psi_n(z) \rangle ~,
\label{avpot}
\ee
where in the latter expression the result of explicit calculation of the
integral over $\vec x$ of the expression from Eq.(\ref{pot1}) is used. The
latter average corresponds to the condition (\ref{norm}) for {\it any}
eigenstate\footnote{In the two-dimensional analog of the $z$ equation
(\ref{zeig}) this is the virial theorem for harmonic oscillator: the average
potential energy is one half of the total.} if the averaged in the last expression in Eq.(\ref{avpot}) function of $z$ is
$4 \, (z^2+S-1)$. Taking into account also the factor $-C$ in the Hamiltonian
(\ref{hw}) one can finally write
\be
V(z, \vec x)  = -c \,(z^2+S-1) \, z^4 \, \int_0^\infty \, {\rm d} \tau \, \left ( {1 \over 2
\pi \tau} \right )^{3/2} \, \exp \left(-  {x^2 \over 2\tau} \right ) \, {
e^{-\tau} \over \sinh^3 \tau} \, \exp \left ( - {z^2 \over 2}\, {e^{- \tau}
\over  \sinh \tau} \right )
\label{potf}
\ee
with $c=C \, \sigma^{3/2}$ being the dimensionless value of $C$.

\section{Binding of excited mesons to heavy quarkonium}
The problem of existence of bound states between the heavy quarkonium and light
mesons is thus reduced to the problem of existence of localized in $\vec x$
solutions of the eigenvalue equation (\ref{5deqm}). General physical arguments
suggest\,\cite{dv} that such solutions exist at a given $c$ and sufficiently
large
excitation in $n$ or/and $S$. In practice a full analysis in $z$ and $x$ of
states excited in $n$ is significantly complicated by the presence of multitude
of lower energy states with lower $n$ and nonzero spatial momentum $p$. For this
reason we analyze here in some detail the excitation of the light meson in the
spin $S$ while keeping $n=0$. The treatment of such excitations is greatly
simplified due to the fact that bound state in question is the ground state in
the `radial' equation (\ref{5deqm})  at a given $S$, so that the corresponding
eigenfunction $\chi_0(z,x)$ has no zeros. We thus use a straightforward
variational procedure to establish existence of bound states at any nonzero $c$,
provided that $S$ is sufficiently large, and then find numerically the onset of
the binding as a function of the coefficient $c$ at few low values of $S$.

For a variational treatment of the binding problem we choose the {\it Ansatz}
for the eigenfunction in a factorized form
\be
\chi_0(z,x)= z^{S+{1 \over 2}} \, e^{- z^2/2} \, \xi(x)~,
\label{ansatz}
\ee
where the $z$ dependence is that of the ground-state wave function of the $z$
equation (\ref{zeig}) at a given $S$. Substituting this variational function in
Eq.(\ref{5deqm}) one finds that the upper bound for the energy of the ground
state of the system is then given by the eigenvalue $\omega$ found from the $x$
equation
\be
[-\Delta_x + U(x)] \, \xi(x) = (\omega^2-4\,S) \, \xi(x)~,
\label{varxi}
\ee
with the effective potential
\begin{eqnarray}
\label{uofx}
&&U(x) = \\ \nonumber
&&-8 \, c \, (S+1) \, (S+2)  \int_0^\infty {\rm d}\tau \, 
{ (S+3)\, \left ( 1- e^{ - 2 \, \tau  } \right ) + S-1  \over (2 \pi \tau)^{3/2}} \, \exp \left(-  {x^2 \over 2\tau} \right )
  \, \left ( 1- e^{ - 2 \, \tau  } \right )^{S} \, e^{- 4 \, \tau}~.
\end{eqnarray}
Considering large $S$ one can readily see that the factor $( 1- e^{ - 2 \, \tau 
}  )^{S}$ effectively cuts off the contribution of values of $\tau$ above 
$\tau_0 \propto 1/\ln S$, so that the range of the potential behaves as $X_0 
\propto 1/\sqrt{\ln S}$. On the other hand the integral over $\vec x$ of the 
potential is proportional to $1/S$, which is weaker than the overal factor 
$\propto S^3$ in front of and in  the integral in Eq.(\ref{uofx}). Therefore a 
bound state exists at sufficiently large $S$ for any nonzero value of the 
coupling strength $c$, which is necessarily positive, due to the positivity of 
the chromo-polarizability.

\begin{figure}[ht]
\begin{center}
 \leavevmode
    \epsfxsize=16cm
    \epsfbox{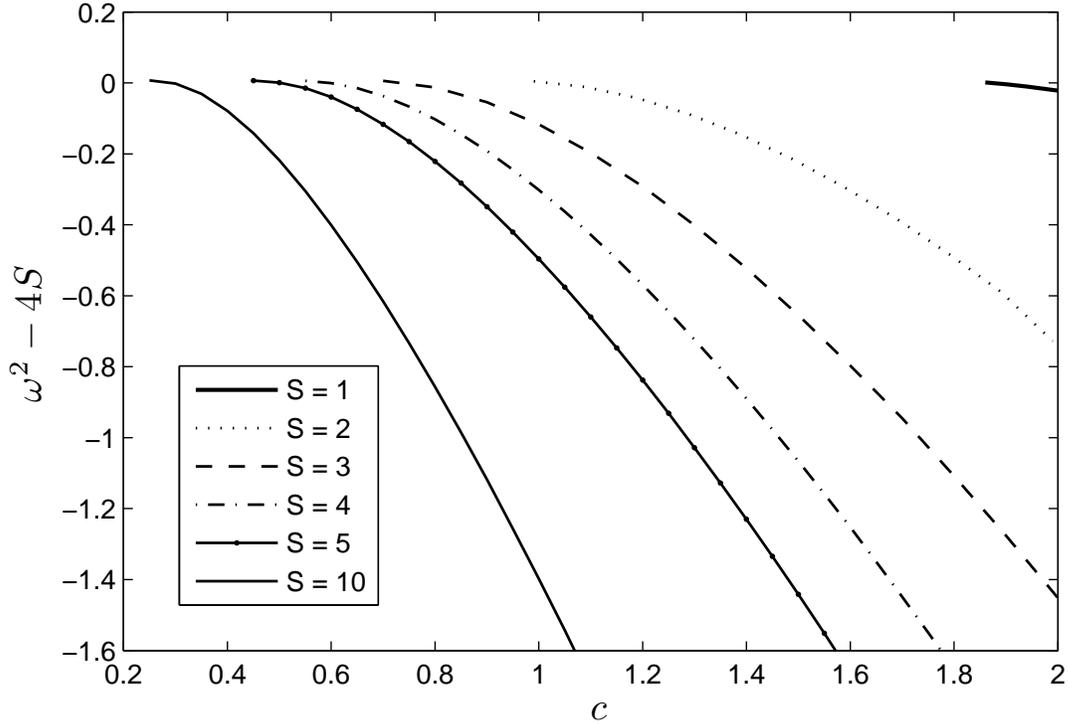}
    \caption{The values of the ground state `binding energy' $\omega^2 - 4 \, S$
as a function of the coefficient $c$ in Eq.(\ref{potf}) for few values of $S$. }
\end{center}
\end{figure}

The results of such variational treatment of the problem of binding are
confirmed by an explicit numerical solution of the two dimensional (in $z$ and
$|\vec x|$) Schr\"odinger equation with the potential (\ref{potf}). For a fixed
$S$ we find that a localized in $x$ ground state solution appears starting from
a critical ($S$ dependent) value of $c$. The behavior of the `binding energy' $
\omega^2 - 4 \, S$ as a function of $c$ is shown in Fig.1 for few values of $S$.
One can readily see that the critical value of $c$ decreases with $S$.

Phenomenological applicability of our conclusions to actual charmonium or
bottomonium critically depends on the values of the chromo-polarizability for
specific quarkonium resonances. Using the estimate for the scale factor $\sigma$
as\,\cite{kkss} $\sigma \approx m_\rho^2/4 \approx 0.15\,$GeV$^2$ we find for
the
`reference' value of $\alpha^{\jp \, \psi'}$: $c \approx 1.0$.  According to the data presented in Fig.1 such value of $c$ would correspond to the appearance of binding at $S > 2$. However, as already mentioned, the diagonal chromo-polarizability of the $\pp$ resonance is likely to be somewhat larger, so that bound hadro-charmonium states may exist already at $S = 2$. 

\section{Hadro-quarkonium decay into pairs of heavy flavored mesons}
The notion of hadro-quarkonium necessarily assumes that the heavy quark and
antiquark stay bound together as quarkonium inside the host light-matter
resonance. In other words the forces from the light quarks do not split the
heavy quarkonium, which would result in the decay of the whole system into final
states with pairs of heavy flavored mesons, which decays are conspicuously
suppressed\,\cite{noother} for the discussed $Y$ and $Z$ resonances. The problem
of dissociation of the heavy quarkonium can in fact be treated, to an extent, in
the limit of large mass $M_Q$, in a semiclassical approximation.

Generally the problem can be formulated as that of a reconnection of strings in
terms of a holographic correspondence, using, instead of the braneless soft wall
model, the underlying scheme with the flavor branes localized in the radial
coordinate (for a recent review see e.g. Ref.\,\cite{fb}). The  position of  the
flavor brane in the radial direction is fixed by the quark mass of the
corresponding flavor. The mesons in this approach correspond to the open strings
with ends on the flavor branes. The ends can be on the same flavor brane, in
which case the meson involves quarks of one flavor, or on the different flavor
branes which corresponds  to a meson built from different types of quarks. Any
such model involves a horizon in the radial coordinate which reflects the
emergent QCD scale in the holographic picture. The flavor brane corresponding to
the light meson is close to the horizon while the brane corresponding to the
heavy flavor is far from it.

A bound state of hadro-quarkonium thus involves two strings. One string with the
ends on the light-flavor brane(s) and the other on the heavy-flavor one. The
decay into states with open heavy flavor mesons corresponds to the reconnection
of the strings into the configuration with two open strings both connecting the
heavy-flavor brane with the light-flavor one(s). This process is somewhat
similar to the previously considered decay of a single excited
meson\,\cite{son2}
via a reconnection of open string ends, which has been interpreted as a
Schwinger type pair production and treated semiclassically. The discussed here
reconnection of the strings between light and heavy flavor branes can similarly
be traced to an {\it induced} production of heavy quark pair near the threshold,
which approach will be presented elsewhere, and which can be useful in a more
elaborate than here treatment.

The leading at large $M_Q$ behavior of the probability of heavy quarkonium
dissociation can be evaluated by a simple nonrelativistic quantum-mechanical
consideration, using a potential model for quarkonium combined with a stringy
picture at larger separation between quarks. Indeed a potential $V(r)$ between
heavy quark and antiquark in quarkonium generically is assumed to be of the
form\,\cite{egmr,mvc}:
\be
V_Q(r) = V_c(r) + \sigma \, r~,
\label{potmod}
\ee 
where the linear term is associated with the stringy behavior at long distances,
while the term $V_c$ is the short-distance `perturbative' part, e.g. in the
one-gluon exchange picture this is a Coulomb-like potential $-a/r$. In the
hadro-quarkonium configuration the total energy of the string is $\sigma \, (R+
r)$, where $r$ is the distance between the heavy quark and the antiquark, and
$R$ is the length of the light-meson string. When the string reconnects between
heavy and light quarks, there is no string at all between the heavy quark and
antiquark, so that the minimal energy of the reconnected configuration is
$\sigma \, (R - r)$. Therefore the dependence of the minimal (over the relative
orientations) energy on $r$ becomes in fact given by
\be
V_{HQ}(r) = V_c(r) - \sigma \, r~,
\label{vhq}
\ee
and this effective potential presents a barrier, tunneling through which results
in dissociation of the heavy quarkonium. The tunneling rate then can be
estimated as
\be
\Gamma \propto \exp \left ( -2 \, \int |p| \, {\rm d} r \right ) \sim \exp \left
( -A \sqrt{M_Q \over \Lambda_{QCD}} \, \right )~,
\label{rate}
\ee
where the numerical constant $A$ critically depends on the presently unknown
position of the energy of the quarkonium below the top of the barrier. In a
generic case of this energy gap being of order $\Lambda_{QCD}$ this coefficient
is parametrically of order one, and the rate is strongly suppressed for heavy
quarks.

\section{Summary}
To summarize. We have considered the van der Waals type interaction between a
heavy quarkonium and an excited light meson within a soft wall holographic model
of QCD. In this model the heavy quarkonium acts as a source of the dilaton field
localized at the boundary. The interaction of the light matter with the dilaton
field is then described by a 5D equation with the potential derived from the
dilaton propagation from the boundary to the bulk. We have shown by a
variational estimate that such interaction necessarily results in bound
hadro-quarkonium states localized near the source at sufficiently high spin
excitation of the light meson. We have also found by a numerical calculation
that bound states arise at low values of the spin already for moderate values of
the strength of the source, which values are close to those expected for
charmonium phenomenology. Furthermore, it is argued that the hadro-quarkonium
resonances are metastable with respect to dissociation into final states with
open-flavor heavy mesons in the limit of large heavy quark mass. The behavior of
the hadro-quarkonium states argued in the present paper agrees well with the
observed properties of some of the Y and Z resonances near the open charm
threshold.

\section*{Acknowledgments}
The work of S.D. and M.B.V. is supported in part by the DOE grant
DE-FG02-94ER40823. The work of A.G. is supported in part by the grants
RFBR-06-02-17382, INTAS-1000008-7865, PICS-07-0292165. A.G. acknowledges the
warm hospitality at the William I. Fine Theoretical Physics Institute where this
work was done.

\end{document}